\documentclass{llncs}
\usepackage{graphicx}
\usepackage{amsmath}
\usepackage{listings}
\usepackage{amssymb}
\usepackage{fancyvrb}
\begin{document}
\title{Combining Static and Dynamic Analysis for Vulnerability Detection}
\author{Sanjay Rawat \and Dumitru Ceara \and Laurent Mounier \and Marie-Laure Potet}
\institute{VERIMAG laboratory\\ University of Grenoble (UJF)\\ 38610 Gi\`{e}res, France.\\
\email{\{Dumitru.Ceara, Laurent.Mounier, Marie-Laure.Potet, Sanjay.Rawat\}@imag.fr}}


\maketitle

\begin{abstract}
In this paper, we present a hybrid approach for buffer overflow detection in C code. The approach makes use of static and dynamic analysis of the application under investigation. The static part consists in calculating \emph{taint dependency sequences} (TDS) between user controlled inputs and vulnerable statements. This process is akin to program slice of interest to calculate tainted data- and control-flow path which exhibits the dependence between tainted program inputs and vulnerable statements in the code. The dynamic part consists of executing the program along TDSs to trigger the vulnerability by generating suitable inputs. We use genetic algorithm to generate inputs. We propose a fitness function that approximates the program behavior (control flow) based on the frequencies of the statements along TDSs. This runtime aspect makes the approach faster and accurate. We provide experimental results on the Verisec benchmark to validate our approach.  
\end{abstract}

\section{Introduction}\label{sec:intro}

\subsection{Context}

In a recent report, published by Veracode \cite{veracodeReport10}, it is pointed out that a high percentage of commercial and open source softwares is written in C/C++ and thereby, making them highly susceptible to many dangerous attacks. Under this category of C/C++, buffer overflow (BoF) tops the list with 32\% of the total vulnerabilities. In the present work, we focus on BoF vulnerability in C code. There have been numerous studies in the literature to detect the presence of BoF in applications \cite{cowan_BoFsurvey00}\cite{john_BoFcomparison03}. These studies can broadly be categorized under static and dynamic code analysis approaches. Both of the approaches have their pros and cons. 

Static code analysis is capable of analyzing all the possible paths from the tainted source to potentially vulnerable statements. However, it may produce false positives when dealing with complex language constructs or non trivial sanitization functions. It is also sensitive to the artifacts introduced by the compiler. For example, an \emph{off-by-one} error can be detected by a static source code analyzer, but its actual exploitability will be known only after the compilation. 

On the contrary, Dynamic analysis can be very accurate in vulnerability detection, but it faces a hard time in finding paths that activate the vulnerability. 
The discipline of generating inputs that trigger a vulnerability is termed as fuzzing \cite{miller90}. Fuzzing can be completely random or \emph{intelligent} \cite{suttan_fuzzBook07}\cite{jared_evolveFuzz06}. 
In the later case, a possible approach is to use a {\em symbolic execution technique} to produce a list of constraints that should be satisfied in order to execute a given path (the so-called {\em path-condition}). These path constraints are presented to a constraint solver to get some possible inputs that satisfy them.  However, the obtained results highly depend on the solver's efficiency. 
An alternative approach consists in re-phrasing the input generation as a {\em search 
problem}, that can be solved using dedicated techniques like genetic algorithms  \cite{Mantere_review05}. 

\subsection{Our proposal}

In this paper we propose a hybrid approach to detect BoF vulnerabilities. First, a static code analysis is used to generate \emph{Taint Dependency Sequences} (TDS) representing a subset of tainted paths leading to a potential vulnerability. Then, a dynamic analysis is used to generate concrete inputs allowing to execute one of these paths. This dynamic part relies on a genetic algorithm.
The fitness function that we use associates a score to each current input according to the runtime dynamics of the application: each obtained execution trace is compared with the TDS that we target, and a next generation of inputs is produced from the best individuals. This fitness function  makes use of the so-called \emph{frequency spectrum} of statements introduced in \cite{ball_concept99}. 

\begin{figure}[h]

\centering
\includegraphics[scale=0.30]{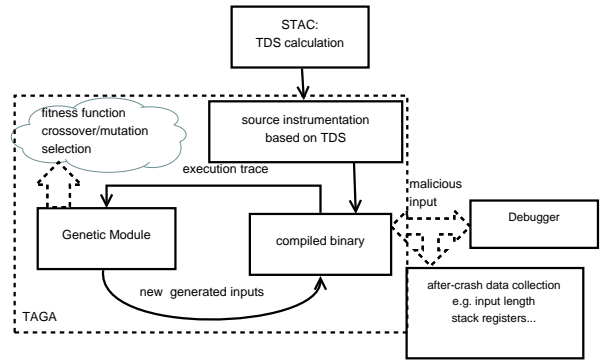}
\caption{A high level flow diagram representing various components of hybrid approach}
\label{fig:tds_ga_high}
\end{figure}

A high level diagram of our approach is depicted in Fig. \ref{fig:tds_ga_high}.
STAC is the static analysis component of the approach. It generates TDSs by performing taint data-flow analysis. Based on this information, the source code is instrumented in order to collect the execution traces produced by running the program. The dynamic component, called TAGA (TDS  Assisted Genetic Algorithm, shown in dotted box), is then used to generate inputs as described above. As we are targeting BoF vulnerability, TAGA keeps generating inputs until we get a program crash, 
or until some threshold is reached in the number of iterations performed by the genetic algorithm. When a crash is obtained, we run the binary with a debugger to collect information about various internal structures (e.g., the execution stack).  This information will provide some useful hints about the {\em severity} of the vulnerability: depending on the content of the execution stack, we can infer how easy it is to write an {\em exploit} for this vulnerability. Thus our approach shows not only the presence of the vulnerability but also its feasibility of exploitation in real world. 

The main contributions of the approach are:
\begin{itemize}
 \item Automated malicious input generation;
\item Robust taint data-flow based execution path generation, thereby reducing search path. This point separates our approach from the ones based on code coverage;
\item New runtime program behavior based fitness function; 
\item Hints on the level of exploitability of a particular vulnerability.
\end{itemize}

The rest of the article is organized as follows. In section \ref{sec:tds}, we briefly describe the static analysis component STAC. We elaborate on TAGA and its various components including fitness function in section \ref{sec:ga}. Section \ref{sec:tool} presents the implementation details and some experiments performed on benchmark data. Section \ref{sec:related} walks over the existing literature to report relevant work and provides discussion in relation to our results. We conclude in section \ref{sec:conclusion} by providing summary and future extension of the approach.

\section{Taint Dependency Sequence}\label{sec:tds}

As quoted in the introduction, one of the main difficulties when using search based 
algorithms for vulnerability detection is to identify the part of the program 
that needs to be covered during the search. In this section, we describe the static 
analysis component that we use in our approach to automate this task.


From a security point of view, a program execution sequence can be considered as ``dangerous'' if it allows 
some user inputs to (directly or indirectly) activate a (potentially) vulnerable statement, like an array access. 
Tracking these dependencies between program inputs and vulnerable statements can be obtained by means of a well-established 
technique called {\em taint analysis}.  

Roughly speaking, taint analysis consists in assigning a {\em taint information} to a variable $v$ at each program
location $l$ if the value of $v$ at this location may depend on an external input.
Consider for instance the example (taken from \cite{verisec08}) shown in Listing.~\ref{fig:cCode} which serves as working example through out the remaining article.
This program takes two string arguments as input, called {\tt gecos} and {\tt login}. It builds a result string ({\tt buf})  by transforming the {\tt gecos}  argument as follows: each occurrence of character ``{\tt \&}'' found before either a ``{\tt ,}'', 
``{\tt ;}'', ``{\tt \%}'' or ``{\tt \verb+\0+}'' character is substituted with the {\tt login} argument.

\lstset{language=[ANSI]C++}
\begin{lstlisting}[basicstyle=\tiny, numberstyle=\tiny, numbers=left, frame=single,name={buildfname simplified vulnerable function},
                   caption=\lstname, captionpos=b, label=fig:cCode]
void buildfname(char *gecos, char *login, char *buf)
{
  char *bp = buf;
  char* p;
  for (p = gecos; *p != '\0' && *p != ',' &&
        *p != ';' && *p != '%'; p++) {
		if (*p == '&') {
			(void) strcpy(bp, login);    /* BAD */
			*bp = toupper(*bp);
			while (*bp != '\0')
				bp++;
   		} else {
			*bp++ = *p;                 /* BAD */
		}
	}
  *bp = '\0';                                      /* BAD */
}
int main(int argc, char **argv)
{
	char *gecos, *login, buf[512];
	gecos=argv[1];
	login=argv[2];
	buildfname(gecos, login, buf);
	return 0;
}
\end{lstlisting}

In this program, parameters {\tt gecos} and {\tt login} are tainted at location 21 and 22 (since they depend on command line arguments).
This taint information is propagated to variables {\tt p} and {\tt bp}, both by {\em data dependency} (through explicit assignments)
and {\em control dependency} (since these variables are assigned within an iteration controlled by an expression depending on {\tt gecos}). 

 In~\cite{dumitru_tds10}, we proposed a static analysis approach which associates
a {\em taint environment} to each variable $v$, at each program location $l$. In addition to taint information, this environment also 
associates to each pair \mbox{($l$, $v$)} a set of TDS explaining {\em why} the variable $v$ 
is tainted at location $l$.  More precisely, each TDS $t = <l_1, l_2, \dots, l_n>$ is a sequence of program locations $l_i$ a program 
execution path should traverse in order to reach $l$ with an input-dependent value assigned to $v$. Thus, when $l$ corresponds to a 
vulnerable statement, this TDS set exactly characterizes the set of ``dangerous'' execution paths. 

In the {\tt buildfname} example depicted in Listing.~\ref{fig:cCode}, 3 vulnerable statements can be identified (lines 8, 13 and 16).
Indeed, in each of these statements a write access to a buffer {\tt buf} is performed on a (potentially) tainted address that might overflow 
the buffer. The TDS sets associated with these locations corresponds to all the execution paths that may activate these vulnerabilities 
(i.e., reach them with a tainted address)\footnote{Since these TDS are computed statically some of these paths may be infeasible.}.

Let us consider, for instance, the vulnerability at line 8. At this location, variable {\tt bp} can be tainted in several ways, e.g.:
\begin{itemize}
\item {\tt *bp} is assigned {\tt p} (at location 13), which was assigned the tainted value {\tt gecos} (location 5).
	The corresponding TDS, therefore, is $<21, 5, 13, 8>$.
\item {\tt bp} is assigned within iteration loops controlled by a tainted value (location 5 and 10).
	Corresponding TDS are $<21, 5, 13, 8>$ or $<22, 8, 10, 11, 8>$ 
\end{itemize}

\section{Genetic Algorithm Based Input Generation}\label{sec:ga}
In this section, we briefly introduce the topic of genetic algorithm used in the TAGA component. Genetic algorithm (GA, for short)  is the member of family of search-based algorithms known as \emph{evolutionary algorithms} \cite{Luke_GAbook09}. GA is based on the idea of human evolution i.e. \emph{survival of the fittest}. Following steps are involved in a typical GA based solution to a problem:
\begin{enumerate}
 \item Based on the problem, an \emph{initial population} of candidate solutions (inputs) is generated randomly.
\item A \emph{fitness function} is defined to evaluate each candidate solution (i.e. its proximity to the target solution).
\item \emph{Crossover} and \emph{mutation} are applied to parents (old generation) to create children (new inputs).
\item A \emph{selection} strategy is applied to generate new population based on fitness score.
\end{enumerate}
 The pseudo-code of GA is given in listing: \ref{fig:GA} (taken from \cite{Luke_GAbook09}). In the following sections, we describe GA components with respect to our implementation. Genetic algorithm is used to generate inputs that cause tainted path (i.e. TDS) to be executed. Indeed, the purpose of a TDS $t = <l_1, l_2, \dots, l_n>$ is to ``guide'' the program execution towards a vulnerability along a path traversing each $l_i$. The idea is to run the vulnerable program with initial inputs and then by using GA generate new inputs that not only traverse a given TDS but also cause the program to misbehave, for example causing crash. 
\lstset{language=[ANSI]C++}
\begin{lstlisting}[basicstyle=\tiny, numberstyle=\tiny, numbers=left, frame=single,name={Pseudo-code for Genetic Algorithm},
                   caption=\lstname, captionpos=b, label=fig:GA]
popsize := desired population size
P := []
for popsize times do
  P := P U <new random individuals>
Best := []
repeat
  for each individual Pi in P do
  AssessFitness(Pi)
  if Best ==[] or Fitness(Pi) > Fitness(Best) then
    Best := Pi
  Q :=[]
  for popsize/2 times do
    Parent Pa := SelectwithReplacement(P)
    Parent Pb := SelectwithReplacement(P)
    Children Ca, Cb := Crossover(Copy(Pa), Copy(Pb))
    Q := P U <Mutate(Ca), Mutate(Cb)>
    P := Q
until Best is the ideal solution OR we have run out of time
return Best
\end{lstlisting}

\subsection{Initial Population}\label{sec:initpop}

As mentioned before, in order to execute a TDS a set of constraints (corresponding to the {\it IF} and {\it WHILE} conditions associated to this TDS) has to be satisfied. When these conditions are {\em character comparisons}, this implies that the (string) inputs should pass these comparisons.  Therefore, based on static analysis of the source code, performed during TDS computation, we collect information on program inputs and corresponding path conditions. 
For the program, shown in Fig. \ref{fig:cCode}, path conditions are in the form of comparisons with particular characters belonging to the set $C=('\&', ',', ';')$. With this information, we construct regular expression to represent the input and thereupon, generate strings that satisfy that regular expression. In our example, we use \Verb+[a-z$

\subsection{Fitness Function}\label{sec:fit}

As aforementioned, our aim is to traverse a given TDS, which may contain nodes responsible for path conditions. We should prefer inputs that traverse through maximum number of such nodes. Based on this criteria, we assign weights to each of the elements belonging to a selected TDS. Our fitness function $F_i$ of $i^{th}$ input ($i \in I$, set of inputs) is defined as follows:
\begin{equation}\label{eq:weight}
 \centering
F_i=\displaystyle\sum_{j=1}^k {w_j}\times f_{ij}
\end{equation}
where $w_j \in W$ corresponds to the weight associated with $l_j \in t$ and $f_{ij}$ is the frequency of $l_j$ for $i \in I$. $W$ is the set of weights for each TDS $t$. Selection of appropriate values for $W$ is of paramount importance for our fitness function to yield good results. For example, if we choose to select equal weights for each $t_j$s, and there is a nested structure in the program such that one (or more) $l_j$ is at inner most nested statement, then it will be executed lesser times, as compared to $l_j$ at outer most statements. As a result, inputs which are reaching till outer statements only will have a high fitness values, thereby giving GA a wrong fitness impression. Therefore, for such situation, we need to assign a higher weight to the inner one as compared to the outer one. To tackle this issue, we have following two approaches to work upon.
\paragraph{Static Approach} While calculating TDS, STAC also learns \textit{IsNested} type of structure among $l_j$s. For example, for any two $l_1, l_2 \in t$, we may learn that $l_1$ and $l_2$ are in a same nested conditional statements structure and $l_1$ lies at upper level than $l_2$ in the hierarchy. Therefore, if we have this information from TDS, we can assign relative weights such that inner $l_j$s have higher weights than outer ones.
\paragraph{Dynamic Approach} A careful runtime analysis reveals a lot about execution trace of the program by means of frequencies corresponding to $l_j$s. In \cite{ball_concept99}\cite{ball_region93}\cite{reps_progProfile97} the authors discuss the dynamic analysis of program by means of \emph{frequency spectrum analysis}. In particular, by observing the frequencies of statements, branching structure can be approximated to some extent. As discussed in above section, it may be noted that rare execution statements will have low frequencies as compared to easily executed ones. It may also be noted that statements lie deep inside a nested structure belong to rare statements. Therefore, frequency spectrum of $l_j$s captures \textit{IsNested} type of structure among $l_j$s. To calculate frequency spectrum, we start with a set of inputs $I:=\langle i_1,i_2,...,i_m\rangle$ to get a frequency matrix $freq = \left( \begin{smallmatrix} \cdots \\ f_{ij} \\ \cdots \end{smallmatrix}\right)$, where $f_{ij}$ is the frequency of $l_j$ for the input $i \in I$. As we are counting frequencies of each $l_j$, a given tds $t$ is simplified further by removing duplicate $l_j$s, keeping only the first occurrence of $l_j$. Based on the above assumption, we calculate weights dynamically by counting global frequencies of $l_j$s and inverting them i.e. for a set of inputs $I$ and the matrix $freq$, the weight $w_j$ of $l_j$ is calculated as follows:
\begin{equation}
 \centering
w_j = \frac{1}{\displaystyle \sum_{i=1}^{m} f_{ij}}
\end{equation}
 We make another observation while calculating weights. Labels in TDS, which are closer to vulnerable statement i.e. the last element of a TDS, should have more weights than the ones which are relatively away from vulnerable statement. This is because any input which is reaching nearer to vulnerable statement has greater chance of reaching the vulnerable statement. Given this, we multiply weights with a function whose value increases as we move from beginning to end in a TDS. One example of such function may be the index of the elements in the TDS i.e. for $l_j \in t$, the function value will be $j$. We denote this function as $ProximVar$. Therefore, considering above formulation in mind, we arrive at the following equation for calculating weights.
 \begin{equation}\label{eq:Dweight}
 \centering
w_j = \frac{1}{\displaystyle \sum_{i=1}^{m} f_{ij}} \times ProximVar
\end{equation}
Substituting the values obtained by using eq. \ref{eq:Dweight} into the eq. \ref{eq:weight}, we get the fitness value of each input. These values are sorted in descending order to select inputs to be used in next population generation.\par In the present study, we use dynamic approach to calculate weights. The static approach of calculating weights is incomplete in the sense that it only provides the relation among various $l_j$s. Nevertheless, we need to use another technique to calculate actual values of weights, for example linear programming \cite{grosso08}. Dynamic approach, on the other hand, uses runtime statistics to calculate weights, without requiring to use another technique. 

\subsection{Selection}\label{sec:select}
We follow \emph{elitism} approach to select inputs. In this approach, we choose best two inputs to be included in the next generation i.e. these two inputs will be competing with their own children in the next generation.

\subsection{Crossover and Mutation}\label{sec:cross}
We perform crossover by combining two inputs (parents) to form two new inputs (children). Based on the problem, we may choose single- or two-point crossover. As our inputs are strings, we concatenate substrings of both the parents by interchanging them with each other. In experimentations, the rate of crossover is set to  100\%.\par Mutation is performed by appending a random (smaller) string which is generated by providing a regular expression of our choice. For our working example, we use \Verb+[ab$&

\section{Implementation and Experimentation}\label{sec:tool}
This section details on implementation of our framework and provides a set of experiments to validate the approach.
\subsection{Implementation}
From the implementation perspective, the present work is composed of two major components- STAC and TAGA. STAC is implemented in Caml, using Frama-C framework. Details of its implementation can be found in \cite{dumitru_tds10}. After the source code is analyzed by STAC, we get all the information that we require to perform dynamic analysis. For a given vulnerable statement, we get a set of corresponding TDSs. We select a TDS which has maximum unique nodes to reach vulnerable statement. The reason for doing so is that smaller TDS may be very trivial or may provide less precise path. \par Dynamic component TAGA is composed of instrumented binary of the program, GA to generate inputs and an interface to communicate with a debugger to get after-crash information. In this study, we choose to work with GDB as our preferred debugger. As TDS includes control flow of the program, we get to know which branch of a conditional statement is executed. This information is used by our fitness function. Therefore, based on a particular TDS for a particular vulnerability path, we instrument lines in source code corresponding to the chosen TDS. For example, for the code given in listing \ref{fig:cCode}, one of the TDSs that we select is $\langle21,5,7,8\rangle$ for the vulnerable statement \texttt{strcpy(bp, login)}. Therefore, we instrument statements after each of these labels- 21, 5, 7, 8. In this way, at runtime, we get execution trace of the program in terms of executed statement's frequency.\par GA is implemented as Python module. Inputs to GA module are instrumented binary and regular expressions to generate inputs. Regular expressions are constructed based on the static analysis of the program which extract information on constraints e.g. character matching in \texttt{IF} statements. Such characters, along with other ASCII characters, are included in regular expression to help GA generating valid inputs faster. GA is run until we get \texttt{SIGSEGV} signal or a predefined iteration threshold is reached (1000 iterations in our experimental setup). We look for \texttt{SIGSEGV} signal because BoF will result in invalid memory reference or segmentation fault. Once we get malicious inputs corresponding to \texttt{SIGSEGV} signal, with the help of another Python module, we run the binary with GDB to get more information which is useful for understanding the impact of analysis on exploit generation (see section \ref{sec:vulnExploit}). Using such information, we can infer as how easy (or difficult) it is to write a real exploit for the vulnerability.
\subsection{Experimentation}
For empirical results, we experiment with Verisec benchmark suite \cite{verisec08}. The suite consists of snippets of open source programs containing BoF vulnerabilities of varied difficulties. For us the level of difficulty is directly proportional to the manipulation done on the tainted input by the program i.e. if there are checks on the input before it reaches the vulnerable statement, it is more difficult to generate such input. From this perspective, we find that Verisec suite has programs as simple as \texttt{bind-> CVE-2001-001-> nslookupComplain-small\_bad.c, gxine-> CVE-2007-0406-> main, Samba-> CVE-2007-0453} etc., wherein generating a larger \emph{random} string overflows the buffer, to as hard as \texttt{sendmail-> CVE-2003-0681-> buildfname}, \texttt{edbrowse-> CVE-2006-6909-> ftpls->\\ strchr\_bad.c} etc. wherein input should contain (or does not contain) specific characters (at specific positions) in order to reach the vulnerable statement. 

There are many programs which contain BoF vulnerability mainly due to \emph{off-by-one} error. We could not experiment with such programs as gcc 4.4.1 compiles the binaries such that vulnerable buffer does not appear just above the saved \texttt{frame pointer} (ebp). This makes it impossible to overwrite last byte of saved ebp which is the technique to exploit \emph{off-by-one} error \cite{klog_offByone99}. Therefore in practice, there is no \emph{off-by-one} error as such and without the dynamic analysis, this would have been a false positive. 

Table \ref{tab:result} shows the findings of our analysis on three programs which do string (inputs) manipulations before letting it reach the vulnerable statement. In order to evaluate the gain by including TDS with GA, we compare our results with two other approaches- random fuzzing approach and code coverage based GA approach, similar to \cite{Grosso_GAbof04}. However, even for random approach, we use the same regular expression that we use in our proposed approach to generate the random inputs i.e. random fuzzing is not completely random. For the second approach, we use \emph{gcov} tool for calculating code-coverage. In this case, the fitness function depends on percentage of code covered by an input and number of times a vulnerable statement is executed (and therefore, reached). We do not use any specific path (slicing) to reach vulnerable statement. The weights are selected as per heuristics discussed in \cite{Grosso_GAbof04}.

\begin{table}[h]
\caption{List of the programs used in experimentation}
\scriptsize
\begin{tabular}{p{0.5cm}|p{1.6cm}|p{1.9cm}|p{0.6cm}|p{1.8cm}|p{1.0cm}|p{1.2cm}|p{1.2cm}}
\hline S.No. & Application & Name & \# LoC & Constraints & TDS + GA & coverage + GA & Random inputs \\
\hline 1 & sendmail & mime\_fromqp & 65 &'=n' & 20 & 26 &243\\
       2 & sendmail & buildfname & 52 &'\&' and not(,;\%)  & 6 & 10 & 34\\
       3 & edbrowse & ftpls & 49 & '-{-} ' (in the beginning) & 35 & * & * \\
\hline
 \end{tabular}
\label{tab:result}
\end{table}

In the table, columns 2--3 denote the path of the vulnerable program in the Verisec suite. \emph{lines of code} parameter is given in column 4. Column 5 shows constraints that an input must satisfy in order to reach vulnerable statement. Column 6 shows the number of iterations (generations) taken by TDS based GA to generate inputs that crash the application. These numbers are average taken over 20 different runs of GA. Column 7 shows the same for coverage based GA. Last column shows the number of iteration taken by random fuzzing approach. A noticeable difference (shown as asterisk) comes out in the case of \emph{edbrowse} program. In this case, out of 20 times, coverage based GA could generate malicious inputs only 3 times and random fuzzing could generate malicious inputs only 2 times. This comparison shows the effectiveness of GA enabled inputs generation \emph{viz-a-viz} randomly generated inputs, specially in the case, when we have knowledge about the precise path to reach the vulnerable statement by means of TDS. The experiments also show that even for the small programs, data- and control-flow assisted GA outperforms other similar approaches\footnote{For larger programs, the cyclomatic complexity will be high which will further widen the gap.}.

\subsection{Vulnerability Exploitability}\label{sec:vulnExploit}
Next step in vulnerability analysis is to check if the vulnerability is exploitable in real world by generating exploits. Generating exploits for a stack BoF vulnerability involves getting information about the execution stack when the buffer overflow occurs. When we get a \texttt{SIGSEGV} signal on a particular input, another python module is used to run the program  with GDB to collect after-the-crash information which includes various stack register contents e.g. \emph{return pointer, stack pointer, frame pointer} etc., offset of input that caused overflowing \emph{saved return address}. The purpose of getting this information is to further validate the exploitability of the vulnerability (explained below). Following we show a typical output of our tool for the program at S. No. 3 in the table. In that program, there exists a BoF when \texttt{strcpy()} is called with a fixed length buffer \texttt{user[USERSZ]} and user controlled input. In order to reach the \texttt{strcpy()} statement, the input string must have '\texttt{-- }' as first three characters. For convenience, we denote \texttt{<space>} by \texttt{S} in the output shown in figure \ref{fig:out}.
\begin{figure}[h]
 \begin{Verbatim}[frame=single,numbers=left,fontsize=\scriptsize]
Generation# 35
Malicious inputs: --S1--*a%*S--l42f4*cS8SnSaaSaSS%*S1-*%%Sn%1S
n1%a-nn*1*nn1a-S%%1-*a1*a*a-a*a-%a-%*Sa1S%1a*nn1--%*l&c%al*cgf=
Lengths:  107
Calling GDB...Returned from GDB..
....
EIP is overwritten by:  a*a-  at index:  68
EBP is overwritten by:  *a1*  at index:  64
ESP is pointing to: a*a-%a-%*Sa1S%1a*nn1--%*l&c%al*cg at index:72
\end{Verbatim}
\caption{Output of the tool.}\label{fig:out}
\end{figure}
Line 2-3 shows the malicious input that caused the program to crash, followed by its length 107. Line \# 7 shows the status of \texttt{eip} which is overwritten by \texttt{a*a-} at an offset 68 in the string and line 9 shows the contents pointed by \texttt{esp} at offset 72. With this information, one can construct a real exploit with the following skeleton:
\begin{Verbatim}[fontsize=\small]
<--S...Ax67...><4 bytes address to 'jump esp' instruction>
<...shellcode..(starting at 72th byte>
\end{Verbatim}
Based on this information, we can infer that it is easy to exploit \texttt{edbrowse} program. On the other hand, in the case of \texttt{ mime\_fromqp}, we find it difficult to exploit as values of \texttt{eip, esp} were not always affected \emph{meaningfully} by the user controlled input which makes it hard to construct an  exploit.

\section{Related Work}\label{sec:related}

Application code analysis is gaining importance, as it can help in writing safe code during the development phase by detecting bugs that may lead to vulnerabilities. As a result, tremendous research on code analysis has been carried out by industry and academia and there exist many commercial and open source tools and approaches for this purpose.

\paragraph{Taint Analysis.} The notion of {\em taint variable} has been introduced within the {\sc Perl} language,
with the use of a special execution mode called ``taint mode''. This  idea of computing variable's taintness {\em at runtime}
has been generalized into several tools, see~\cite{Chang08} for a more complete survey.
Input dependencies of program variables can also be computed using static analysis techniques, as in \cite{Wass07,Scholz08,TAJ09}. The main advantage of the latter approach is to consider the whole set of program executions, but, as usual, the price 
to pay is larger number of {\em false positives}. Moreover, it is highly desirable to extend results 
on variable taintness with information on the corresponding execution paths. This point has been considered so far only in a few 
existing works~\cite{Le07,Nagy09}. The most dangerous paths are then used as ``test objective'' in order to check if the corresponding 
vulnerability is effective or not. On similar lines, in \cite{ganesh_taint09}, Ganesh \emph{et al.} proposed an automated white-box fuzzing techniques in the form of a tool- \emph{BuzzFuzz}. The technique is to identify vulnerable points in the code and execute the instrumented binary with \emph{valid} input set; using data taint analysis, identify inputs that affect those vulnerable points; fuzz these inputs and run the binary with them. In this way, the \emph{legal} syntactic structure of the program is maintained while executing it with new fuzzed inputs. However, a domain expert knowledge is required to fuzz the relevant inputs manually.
 
\paragraph{Concolic Approaches.} Concolic approach is a \emph{portmanteau} of concrete and symbolic execution based approaches for analysis. A pure symbolic approach- EXE is proposed in \cite{cadar_exe06}, wherein the program is run on symbolic inputs while recording input-constraint pairs. When a particular path ends or a bug is found, EXE generate a real input by solving the constraints by using a solver. The problem with such approaches is the path explosion and imperfect symbolic execution \cite{Godefroid_FuzzTesting08}. Godefroid \emph{et al.} \cite{Godefroid_FuzzTesting08} implement a tool- SAGE which incorporates techniques from symbolic and dynamic execution. At algorithm level, the approach has components similar to our approach, specially from GA implementation standpoint. However, the functionality of these components is different. The idea is to start with an initial input and capture constraints by executing the program. These constraints are given to constraint solver. Based on this solution and their \emph{score} in covering basic-blocks, new children are generated to cover new paths. This process is continued until some interesting event is captured. The analysis is performed on machine-code, which makes it suitable for wide variety of compilers, compiling to x86 instructions. 

\paragraph{GA Assisted Dynamic Approaches.} Dynamic analysis technique analyzes the application by executing it on real inputs. Miller \emph{et al.} coined the term \emph{fuzzing} to test UNIX utilities by generating random inputs to discover malfunctioning \cite{miller90}. The idea is to generate random values as inputs that covers boundary cases. But as noted in \cite{cadar_exe06}\cite{ganesh_taint09}, vulnerabilities that are buried deep inside the code are hard to discover using this technique. Genetic algorithms (GAs) have been used to generate test cases for program testing \cite{wasif_review09}\cite{Mantere_review05}. Grosso \emph{et al.} have used GA to generate inputs automatically that trigger BoF vulnerability in the application \cite{grosso08}\cite{grosso05}. The fitness score of input is based on its ability to cover code and reaching vulnerable statements and weights associated with each of these factors. They have used linear programming to adjust weights automatically which is an additional step in the whole process of input generation. The authors mention to consider \emph{nested} statements, but have not provided enough description of its detection in source code and usage in the fitness calculation. In our approach, we have made use of readily available statement frequencies to approximate nested statements and their weights. Sidewinder is a tool for analyzing binaries to detect vulnerabilities using GA assisted fuzzer \cite{sparks_sidewinder07}. In this approach, control flow is modeled as Markov Process and fitness function is defined over Markov probabilities associated with state transition on control flow graph. Inputs are generated using \emph{grammatical evolution}. Due to the dependency of fitness function on transition probabilities, this approach should be able to reach \emph{deeply nested} statements which is important if vulnerable statements are buried deep inside. Our approach is closest to this approach with added feature of simplifying the weight calculation and path to traverse to reach vulnerable statements. On the similar lines, Liu \emph{et al.} construct \emph{control dependence predicate path} (CDPPath) from the binary of the application and apply GA to construct inputs to reach vulnerable statements \cite{liu_x86GA08}. Their fitness function depends on the number of predicates in CDPPath covered by inputs. However, this study treats each predicate equal which may result in stagnation during later stage of searching.

\section{Conclusions and Future Work}\label{sec:conclusion}
In this article, we report our work of generating inputs to reach a vulnerable statement and putting the program in an abnormal states. Static analysis based taint data-flow approach has the advantage of producing all possible paths from tainted inputs to vulnerable statements, but is generally blind towards paths constraints that must be satisfied by an input to traverse a given path. Symbolic execution tries to overcome this hurdle by solving path constraints to generate inputs, but at the cost of high computational complexity. Random fuzzing is surprisingly effective in generating inputs for corner cases in lesser time. This approach, however, does not consider any conditions on the inputs, thereby making it unsuitable for more complex programs. We, therefore, propose a hybrid approach by combining static and dynamic analysis to extract nicer properties of both the approaches. By doing a static analysis, we extract TDS- a program slice from tainted input to vulnerable statements and string constraints to be satisfied by the inputs. Rather than solving these constraints, we use GA to generate inputs along a TDS. The fitness function depends on the runtime behavior of the program in terms of execution frequencies of statements. This amalgamation makes the approach faster and accurate, which is further supported by empirical results that we obtain on a benchmark dataset.\par This paper reports preliminary results of our work. We envisage many improvements that comprise our future work. We intend to provide more formal makeup to \emph{frequency spectrum analysis}. Another point of difficulty is that when more conditions are checked in a single $IF$ statement, GA has no way to knowing up to what percentage of predicates are being satisfied by an input. This makes it difficult to select inputs for its next generation. Had it been represented by multiple separate $IF$ statements, GA can observe the respective frequencies of separate $IF$ statements to select inputs. It should be noted that at binary level, this structure is available by default. Each condition in a composite $IF$ statement is represented by a separate basic-block. This give us  incentive to shift our static analysis to binary code. We intend to integrate it with a debugger to have better runtime control. Finally, for a better coverage of all tainted paths to a vulnerability, we plan to design a fitness function (similar to the one proposed in \cite{cao_multipath09}) which is based on more than one TDS corresponding to one particular vulnerable statement.      

\bibliography{tdsGA}
\bibliographystyle{splncs03}
\end{document}